\documentstyle[prl,aps,times,twocolumn,epsfig,floats]{revtex}

\begin{document}

\sloppy

\draft

\title{Screening of Hydrodynamic Interactions in \\
  Semidilute Polymer Solutions: A Computer Simulation Study}

\author{Patrick Ahlrichs, Ralf Everaers and Burkhard D\"unweg}

\address{Max--Planck--Institut f\"ur Polymerforschung,
  Postfach 3148, D--55021 Mainz, Germany}

\address{
  \begin{minipage}{5.55in}
    \begin{abstract} \hskip 0.15in
      We study single--chain motion in semidilute solutions of polymers of
      length $N = 1000$ with excluded--volume and hydrodynamic interactions by
      a novel algorithm. The crossover length of the transition {from} Zimm
      (short lengths and times) to Rouse dynamics (larger scales) is
      proportional to the static screening length. The crossover time is the
      corresponding Zimm time. Our data indicate Zimm behavior at large
      lengths but short times. There is no hydrodynamic screening until the
      chains feel constraints, after which they resist the flow: ``Incomplete
      screening'' occurs in the time domain.
    \end{abstract}
    \pacs{PACS Numbers: 83.10.Nn, 05.40.Jc, 47.11.+j, 61.25.Hq, 36.20.Ey}
  \end{minipage}  
  }

\maketitle

The dynamics of polymer chains in solution
\cite{doi:86,deGennes:79}
has been the subject of long--standing theoretical investigations, even for
the simple case of flexible uncharged chains in good solvent. While in the
dilute limit the validity of Zimm scaling predictions
\cite{doi:86,deGennes:79,akcasu:80,oono:85}
is generally accepted, as confirmed by experiments
\cite{akcasu:80,martin:86,bhatt:89}
and computer simulation
\cite{pierleoni:92,duenweg:93,ahlrichs:99},
the theoretical understanding becomes much more involved as soon as one
considers finite concentrations
\cite{freed:74,degennes:76,freed:81,edwards:84,richter:84,shiwa:88,%
frederickson:90,colby:94}.
This is so due to the complicated interplay between excluded--volume
interactions, hydrodynamic interactions, and entanglements. Only for the
opposite limit of dense melts, where the first two interactions are fully
screened, there exists a well--controlled description in terms of the Rouse or
reptation model \cite{doi:86,kremer:90,ewenreview}. However, the details of
the crossover, the underlying mechanism of the screening of hydrodynamic
interactions, and the concentration dependence of the screening length have
been a subject of considerable debate.

In this Letter, we present the first computer simulation study which is able
to contribute to the resolution of these questions. Experiments, such as light
scattering
\cite{wiltzius:84,wiltzius:86}
or non--equilibrium methods
\cite{zhang:99}
usually focus on {\em collective} concentration fluctuations, while {\em
  single--chain} motions are only accessible by labeling techniques (neutron
\cite{richter:84}
or light
\cite{martin:86}
scattering). Computer simulations can in principle analyze both types of
motion; however, for reasons of statistical accuracy we had to confine
ourselves to single--chain motion.

We study the equilibrium fluctuations of a three--dimensional semidilute
system of flexible bead--spring polymer chains with full excluded volume
interactions, coupled to a hydrodynamic background to fully take into account
hydrodynamic interactions, using an efficient method which we have recently
developed and tested
\cite{ahlrichs:99}.
The polymer system is simulated by Langevin stochastic dynamics, the solvent
by a stochastic D3Q18 lattice Boltzmann model
\cite{ladd:94,ladd2:94},
and a point--particle coupling is introduced via a monomeric friction
coefficient. The present work uses the same model with the same parameters but
in the semidilute regime. One particular advantage, without which the study
would have been unfeasible, is the fact that the lattice Boltzmann solvent
does not alter the good solvent statistics of the chain conformations. We
therefore first equilibrated the multi--chain system without the
computationally expensive solvent, using a combination of stochastic dynamics
and slithering--snake Monte Carlo moves comprising several momomers (roughly
one blob, see below). This run produced a set of configurations, which were
afterwards coupled to the solvent. For further details of the model, we refer
the reader to Ref.
\cite{ahlrichs:99}.

Semidilute systems are characterized by a very low monomer concentration $c$,
which is nevertheless large enough to induce strong overlap of the coils. The
static conformations are well understood
\cite{deGennes:79}
in terms of the ``blob size'' $\xi_S$, i.~e. the typical mesh size of the
temporary network. On scales below $\xi_S$, the chains are self--avoiding
walks (SAWs) characterized by the scaling law $R \sim a N^\nu$, where $a$ is
the monomer size, $R$ the chain extension, $N$ the degree of polymerization
and $\nu \approx 0.59$. The concentration dependence of $\xi_S$ results {from}
$c \sim \xi_S^{-3} (\xi_S / a)^{1/\nu}$. On scales above $\xi_S$ the
excluded--volume interaction is screened, and the chains are random walks
(RWs, $R \sim a N^{1/2}$). The overall chain is thus a RW of blobs with $R^2
\sim \xi_S^2 N / (\xi_S / a)^{1/\nu}$. This picture implies that rather long
chains are necessary in order to clearly observe both regimes; guided by the
idea of having roughly $30$ blobs of $30$ monomers each available, we chose $N
= 1000$, and varied $\xi_S$ by studying the concentration values $c = 0.00837,
0.0397, 0.0734, 0.134, 0.231$ for the statics, and the latter three values for
the dynamics (data are always given in the Lennard--Jones unit system of Ref.
\cite{ahlrichs:99}). The number of chains $M = 50$ was kept fixed; this is
large enough to ensure that even the most concentrated system does not exhibit
self--overlap due to the periodic boundary conditions. Our data for the
static chain conformations are in perfect agreement with the blob scaling
picture \cite{futurelongpaper}, as have been those of previous extensive Monte
Carlo simulations \cite{paul:91,mueller:00}.

\begin{figure}
\begin{center}
\epsfig{file=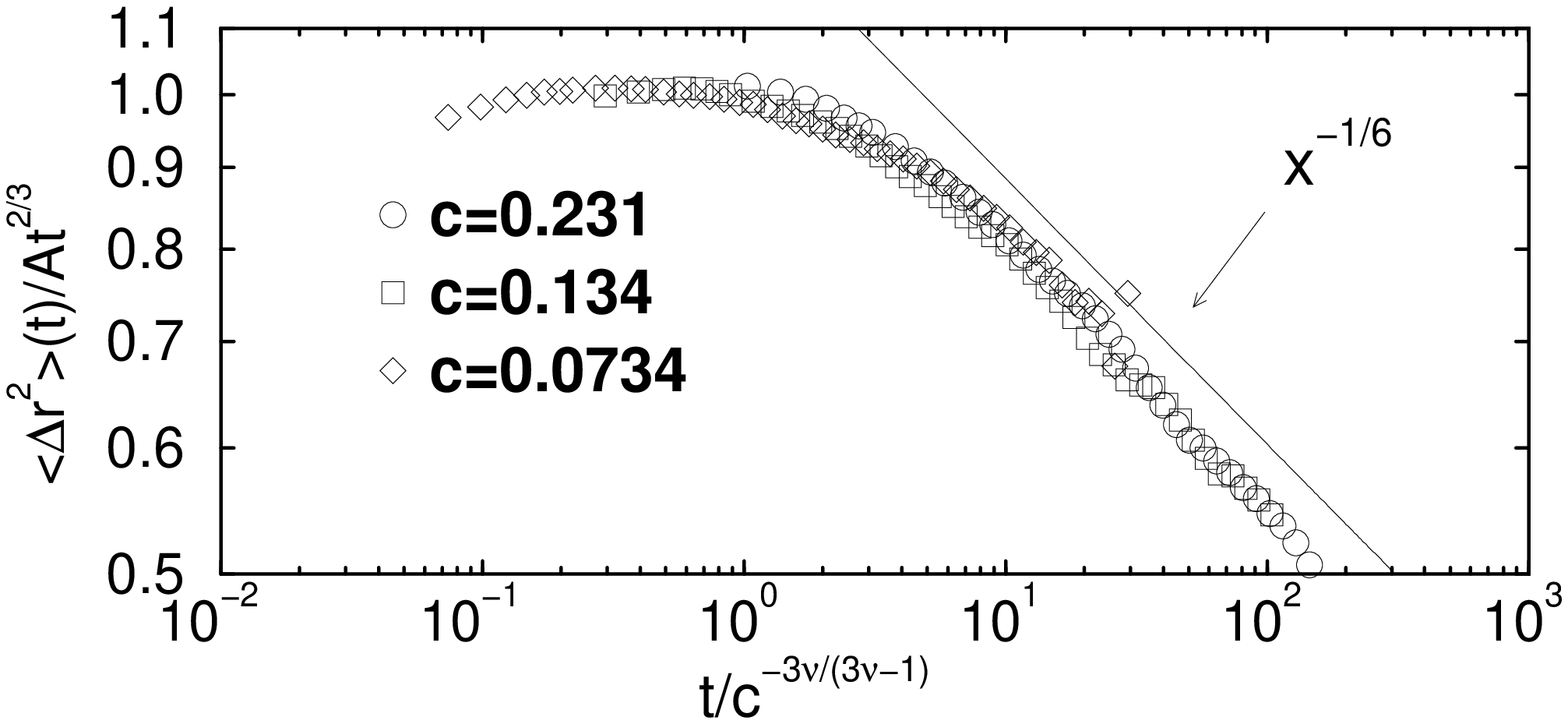,angle=-90,width=7cm}
\end{center}
\caption{Scaled $\left< \Delta r^2 \right>$ for different concentrations.}
\label{fig:g1}
\end{figure}

Dynamic scaling for a single chain which exhibits no special length scale
except $a$ and $R$ implies $\tau \propto R^z$, where $\tau$ is the
conformational relaxation time, and $z = 3$ for the Zimm model (applicable to
dilute solutions without chain overlap where hydrodynamic interactions are
fully developed), while $z = 4$ for the RW Rouse model (applicable to dense
melts where hydrodynamic interactions are fully screened; we do not consider
reptation--like slowing down, which occurs only for longer chains, and does
not play any role for our simulation data). In terms of the chain diffusion
constant $D$ this corresponds, via $D \tau \sim R^2$, to $D \propto R^{-1}$
for Zimm (chain as a Stokes sphere) and $D \propto N^{-1}$ for Rouse (monomers
as independent friction centers). Furthermore, the scaling of lengths with the
corresponding times implies a subdiffusive behavior for the single--monomer
mean square displacement $\left< \Delta r^2 \right> \propto t^{2/z}$, and a
$k^z t$ behavior for the single--chain dynamic structure factor $S(k,t) =
N^{-1} \sum_{ij} \left< \exp \left( i \vec k \cdot (\vec r_i(t) - \vec r_j(0))
  \right) \right>$ in the scaling regime of intermediate length scales
(between $a$ and $R$) and time scales (between $\tau_0$, the microscopic time
for monomer relaxation, and $\tau$).

For a semidilute system, one expects a crossover between these cases. Indeed,
our data for $\left< \Delta r^2 \right>$ do exhibit a crossover {from} a
Zimm--like $t^{2/3}$ behavior at short times to $t^{1/2}$ at longer times. The
behavior at short length and time scales is thus Zimm--like. The pure Zimm
model \cite{doi:86} predicts that the decay rate, i.~e., in the given context,
the prefactor $A$ of the law $\left< \Delta r^2 \right> = A t^{2/3}$ should
only depend on solvent viscosity and not on concentration. We nevertheless
found a weak concentration dependence of $A$ (roughly $20 \%$ within the given
concentration range, see below). Figure \ref{fig:g1} thus studies $\left<
  \Delta r^2 \right> / (A t^{2/3}) = f(t / t_c)$, where $t_c$ is the
concentration--dependent crossover time, which again is the Zimm relaxation
time of a dynamic crossover length (or hydrodynamic screening length), $t_c
\propto \xi_H^3$, and $f(x) \propto x^{-1/6}$ for large $x$. We find a very
good data collapse assuming that $\xi_H \propto \xi_S$ or $t_c \propto c^{- 3
  \nu / (3 \nu - 1)} = c^{-2.3}$, as done in Fig. \ref{fig:g1}. The
assumptions $\xi_H \propto c^{-1}$ \cite{freed:74,edwards:84} and $\xi_H
\propto c^{-1/2}$ \cite{frederickson:90} produced significantly poorer
collapses. The simulation thus confirms de Gennes' prediction $\xi_H \propto
\xi_S$ \cite{degennes:76}, as also observed in experiments (except for weak
corrections \cite{wiltzius:84,zhang:99}).

\begin{figure}
\begin{center}
\epsfig{file=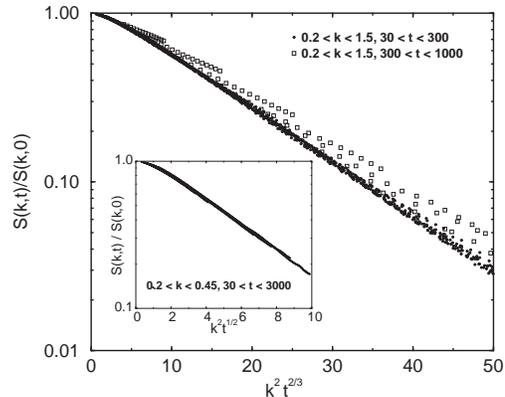,width=7cm}
\end{center}
\caption{Decay of $S(k,t)/S(k,t)$ on length and time scales as indicated,
  suggesting Zimm scaling for short times (main figure) and
  Rouse scaling for long wavelengths (inset).}
\label{fig:scalinginbigranges}
\end{figure}

\begin{figure}
\begin{center}
\epsfig{file=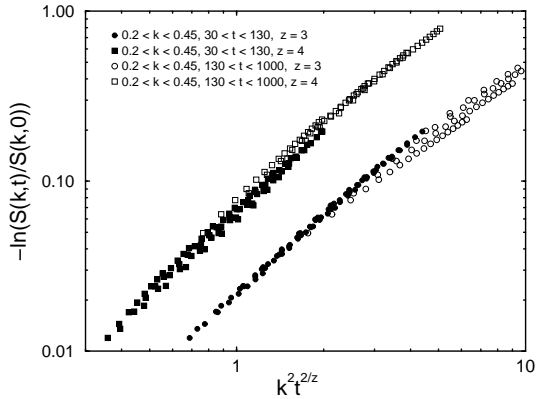,angle=-90,width=7cm}
\end{center}
\caption{$S(k,t)/S(k,0)$ in the RW regime $0.2 < k < 0.45$
  with both Zimm and Rouse scaling, using a representation which emphasizes
  the short--time behavior.}
\label{fig:decideit}
\end{figure}

$S(k,t)$ was studied for the most dilute system $c = 0.0734$. We prefer
scaling plots of the raw data using asymptotic exponents over fits to
functional forms derived {from} approximate theories. {From} $S(k,0)$ we
estimated the crossover wavenumber as $k_c \approx 0.45$ and the scaling
regime as $0.2 < k < 1.5$. {From} $\left< \Delta r^2 \right>$ we read off $t_c
\approx 10^3$; the nonuniversal regime $t < 30$ was discarded. Figures
\ref{fig:scalinginbigranges} and \ref{fig:decideit} show that for short times
$t \ll t_c$ the decay can be described quite well by Zimm scaling, {\em
  regardless of wavenumber}, while for $t \approx t_c$ there is a simultaneous
smooth crossover to Rouse dynamics for those wavenumbers which have not yet
fully decayed, i.~e. for $k < k_c$. Note that the initial Zimm regime of these
wavenumbers can be easily overlooked in the representation of the inset of
Fig. \ref{fig:scalinginbigranges}.

The physical picture which results {from} this observation is thus free Zimm
motion up to the crossover time ({\em on all length scales}), after which
screening sets in, leading to Rouse--like motion. Hence the most important
finding of our simulation is that hydrodynamic screening must necessarily be
viewed as a {\em dynamic time--dependent phenomenon}. We consider this to be
the logical consequence of the (correct) original treatment by de Gennes
\cite{degennes:76} (see below). Nevertheless, this has so far been overlooked
in the literature, the main reason being that single--chain motion on length
scales beyond $\xi$ is not accessible to standard scattering experiments
\cite{wiltzius:84,wiltzius:86} which are sensitive to collective concentration
fluctuations: On scales $k \xi \ll 1$ the overall solution is homogeneous, and
one observes a simple diffusive decay $\exp( - D_{coop} k^2 t)$ with $D_{coop}
\propto \xi^{-1}$. Accordingly, single--chain motion on scales beyond $\xi$
was not treated explicitly in Ref. \cite{degennes:76}. The experiments on
labeled chains \cite{martin:86,richter:84} produced data which are fully
consistent with our view, but were interpreted incorrectly (see below). The
rest of the paper will be devoted to further discussion, and elucidation of
the underlying mechanism.

Hydrodynamic interaction is the presence of long--ranged correlations in the
stochastic displacements of a system of Brownian particles, caused by fast
diffusive momentum transport through the surrounding fluid. They can be
calculated by solving the stationary Stokes equation around a system of
spheres
\cite{mazur:82},
resulting in a complicated multipole expansion which contains many--body terms
representing the multiple scattering of the flow. In the dilute limit,
however, it is sufficient to just consider the leading--order pair
interaction, which decays like $r^{-1}$, where $r$ is the interparticle
distance (Oseen tensor). Conversely, {\em screened} hydrodynamic interactions
are described by a Yukawa--like decay $r^{-1} \exp(-r/\xi_H)$ defining the
hydrodynamic screening length $\xi_H$. Such an interaction occurs for Darcy
flow through a porous medium, where {\em fixed} frictional obstacles with
friction constant $\zeta$ exert a force $- \zeta \vec u$ on the flow with
velocity $\vec u$. Denoting the obstacle concentration with $c$, the flow is
described, on scales beyond the typical interparticle distance, by a modified
Stokes equation $\rho \partial \vec u / \partial t = \eta \nabla^2 \vec u -
\zeta c \vec u$, where $\rho$ is the fluid density and $\eta$ the viscosity.
This implies $\eta \xi_H^{-2} = \zeta c$.

The simplest approach to hydrodynamic screening in polymer solutions just
replaces the $r^{-1}$ Oseen interaction by a screened Yukawa--like
interaction, leading to uncorrelated displacements of monomers whose distance
exceeds $\xi_H$. The resulting motion of the chain is Zimm--like on short
length and time scales and Rouse--like on length scales beyond $\xi_H$ {\em
  for all times} \cite{richter:84}. The Darcy flow thus produces the desired
crossover.

Unfortunately, this picture generates as many questions as it answers. In
particular, the obstacles must be the {\em mobile} polymer chains themselves,
whereas strict Darcy flow requires {\em fixed} obstacles. Moreover, momentum
is present and being transported infinitely far in polymeric as well as in
simple fluids. This fundamental conceptual difficulty was recognized by
Richter {\em et al.}  \cite{richter:84}. In their ``incomplete screening''
model they proposed that the hydrodynamic interaction should cross over to a
second $r^{-1}$ regime on very large scales, but with the {\em overall}
viscosity as a prefactor. For the single--chain short--time behavior this
model also predicts Rouse--like motion. However, this regime is now restricted
to length scales $\xi_H \ll k^{-1} \ll \xi_H \eta_{solution} /
\eta_{solvent}$. On larger scales there is an additional Zimm regime. Richter
{\em et al.}  \cite{richter:84} used this to interpret the mixture of Rouse--
and Zimm--like signals in their scattering data. Similar arguments were used
by Martin \cite{martin:86}, who observed Zimm scaling on all length scales in
the initial decay rate of dynamic light scattering of labeled chains.

The simple model and the more refined version by Richter {\em et al.}
\cite{richter:84} are at variance with both our data and our theoretical
arguments for the short-time behavior, see below. It should be noted that the
``incomplete screening'' model must have fundamental conceptual difficulties,
since $\eta_{solution}$ appears in the short--time dynamics, although
$\eta_{solution}$ is established only on time scales beyond the overall chain
relaxation.

It is therefore clear that a consistent theoretical description has to study
the dynamics of the coupled polymer--solvent system. The first attempt by
Freed and Edwards \cite{freed:74} considered a multiple scattering series of
the flow around the monomers, which is in spirit somewhat similar to the
multipole expansion of Ref. \cite{mazur:82}. After some approximations an
effective Darcy equation results, with $\xi_H \propto c^{-1}$. Later this
scheme was shown to be inadequate; in Ref. \cite{freed:81} it was {\em proven}
that a system of {\em phantom} chains (which do not interact with each other,
but to which the original approach \cite{freed:74} should apply as well) {\em
  does not exhibit any hydrodynamic screening whatsoever}. This is consistent
with new results for colloidal suspensions, where possible hydrodynamic
screening has been discussed recently; the result of large--scale computer
simulations \cite{ladd:96} was the {\em absence} of screening.

With respect to hydrodynamic screening in polymer solutions we can thus draw
the following conclusions: (i) neither does the presence of higher--order
terms of the multipole expansion \cite{mazur:82} at finite concentrations of
scattering centers lead to screening; (ii) nor can such terms be of any
importance in the semidilute limit, where one can reach arbitrarily small
monomer concentrations $c$, while still keeping the chains at strong overlap;
(iii) as this ``colloidal'' mechanism of screening does not apply, the
underlying physics must rather be polymer--specific; and (iv) the mechanism
must lead to a time--delayed screening.

Concerning the short--time behavior, we note that the semidilute system is
governed by a Kirkwood diffusion equation \cite{doi:86}, with a {\em pure
  Oseen--type} $r^{-1}$ diffusion tensor, which describes the short--time
diffusive behavior, and a force term due to connectivity, excluded volume, and
entanglements. Within this formalism, it can be shown rigorously \cite{doi:86}
that the initial decay rates of correlation functions are {\em only} governed
by the diffusion tensor and the statistics of the chain conformations. In
particular, considering the initial decay rate of the single--chain structure
factor \cite{akcasu:80}, one obtains the {\em same} formula as for an isolated
chain in solvent --- the effect of the other chains is merely the modification
of the conformations. Zimm chains, however, have always $z = 3$ independently
of chain statistics; the initial decay rate is given by $\Gamma(k) \sim (k_B T
/ \eta) k^3$, while the fractal dimension only enters the prefactor
\cite{doi:86}. For systems in the semidilute limit we thus conclude, in
accordance with our simulation results and the experimental data by Martin
\cite{martin:86}, that for short times the single--chain dynamics is
Zimm--like, {\em independently of length scales}.

In his pioneering 1976 paper \cite{degennes:76} de Gennes noticed that the
decisive mechanism for screening is the connectivity and the strong coupling
to the temporary matrix (expressed in terms of an elastic gel, which is
physically more appropriate than a rigid porous medium): After the time needed
for a blob to move its own size, which is the blob's Zimm time $t_c$, the
chain will, on average, feel the constraints by the temporary matrix. {From}
then on it is unable to follow the flow, but rather lags behind, and starts to
produce Darcy--type frictional resistance. As the blob can be envisioned as a
Stokes sphere with radius $\xi$ and friction coefficient $\sim \eta \xi$, the
obstacles which produce the Darcy flow are not the monomers but rather the
blobs. Hence, the hydrodynamic screening length is given by $\eta \xi_H^{-2}
\sim \eta \xi_S c_{blob} \sim \eta \xi_S \xi_S^{-3}$, i.~e. the hydrodynamic
screening length is, apart {from} prefactors, {\em identical} to the static
screening length, $\xi_H \propto \xi_S$. This argument \cite{degennes:76}
makes the picture fully self--consistent. On length scales beyond $\xi$, and
time scales beyond $t_c$, the semidilute solution is just a Rouse melt of
blobs, while the conformations within the blobs are already fully relaxed. In
the Rouse regime, momentum transport is no longer described by a simple
Navier--Stokes equation. It rather occurs mainly along the chain backbones,
due to the connectivity forces. This results in a very efficient randomization
of a locally applied ``kick''.

It should be noted that our simulated system deviates somewhat {from} that
ideal scenario. The most dilute system has a density of $9 \%$ of a typical
dense melt, and thus one must expect that higher--order terms in the multipole
expansion \cite{mazur:82} do play a role. We believe that these are the main
source of the $c$ dependence of the prefactor of the initial $t^{2/3}$ law in
$\left< \Delta r^2 \right>$. We expect finite size effects
\cite{duenweg:93,ahlrichs:99}
with respect to the linear box size $L$ to be rather small, since for our data
$(kL)^{-1} < 0.06$.

To summarize, we have presented the first computer simulation study which was
able to study the dynamic crossover {from} Zimm to Rouse behavior in
semidilute polymer solutions. This was made possible by a novel algorithm,
whose essential feature is the replacement of the solvent by a Navier--Stokes
background which is coupled dissipatively to the monomers. Our results are
fully consistent with de Gennes' scaling picture \cite{degennes:76}, and
emphasize the fact that hydrodynamic screening is a dynamic effect which
becomes relevant only after the crossover time. Incomplete screening thus
indeed occurs, however not on large length scales \cite{richter:84}, but on
short time scales. Any theoretical description which builds upon a screened
hydrodynamic interaction depending only on distance, but disregards the time
dependence, cannot describe the phenomena correctly.

We gratefully acknowledge fruitful discussions with K. Kremer, and thank the
Max Planck Society for generous allocation of Cray T3E computer time at RZG.

\end{document}